\documentclass[useAMS,usenatbib]{mn2e}

\usepackage{ graphicx}
\usepackage{ subfigure}

\title[Faint-end Quasar Luminosity Functions from Cosmological Hydrodynamic Simulations]{Faint-end Quasar Luminosity Functions from Cosmological Hydrodynamic Simulations}

\author[Colin Degraf et al.]  {Colin Degraf$^{1}$, Tiziana Di Matteo$^{1}$, Volker Springel$^{2}$\\ $^{1}$ McWilliams Center for
       Cosmology, Carnegie Mellon University, 5000 Forbes Avenue, Pittsburgh,
       PA 15213, USA\\ $^{2}$ Max-Planck-Institut f\"{u}r
       Astrophysik, Karl-Schwarzschild-Stra\ss e 1, 85740 Garching bei
       M\"{u}nchen, Germany\\}

\begin{document}

\date{Accepted 200-.
      Received 200-;
      in original form 200-}

\pagerange{\pageref{firstpage}--\pageref{lastpage}}
\pubyear{200?}
\maketitle

\label{firstpage}
\begin{abstract}
We investigate the predictions for the faint-end quasar luminosity function (QLF) and
its evolution using fully cosmological hydrodynamic simulations which
self-consistently follow star formation, black hole growth and associated
feedback processes. We find remarkably good agreement between predicted and
observed faint end of the optical and X-ray QLFs (the bright end is not
accessible in our simulated volumes) at $z < 2$. At higher redshifts our
simulations tend to overestimate the QLF at the faintest luminosities.  We show
that although the low (high) luminosity ranges of the faint-end QLF are
dominated by low (high) mass black holes, a wide range of black hole masses
still contributes to any given luminosity range. This is consistent with the
complex lightcurves of black holes resulting from the detailed hydrodynamics
followed in the simulations. Consistent with the results on the QLFs, we find
good agreement for the evolution of the comoving number density (in
optical, soft and hard X-ray bands) of AGN for luminosities $\ge 10^{43}$
erg s$^{-1}$. However, the luminosity density evolution from the simulation appears to
imply a peak at higher redshift than constrained from hard X-ray data (but not
in optical). Our predicted excess at the faintest fluxes at $z \ge 2$ does not
lead to an overestimate to the total X-ray background and its contribution is
at most a factor of two larger than the unresolved fraction of the 2-8 keV
background. Even though this could be explained by some yet undetected,
perhaps heavily obscured faint quasar population, we show that our
predictions for the faint sources at high redshifts (which are
dominated by the low mass black holes) in the simulations are likely affected
by resolution effects.

\end{abstract}


\section{Introduction}
In recent years quasars have been used as instrumental tools for probing
properties of their host galaxies as well as large scale structure through
cosmic time.  The existence of black holes at the centre of most galaxies
\citep{1995ARA&A..33..581K} combined with the correlation between supermassive black
holes and their parent galaxies \citep{1998AJ....115.2285M,
2000ApJ...539L...9F, 2000ApJ...539L..13G, 2002ApJ...574..740T,
2007ApJ...655...77G} significantly strengthen the link between the black hole
and the formation and evolution of galaxies. Although the origins of these
correlations are not completely understood, recent observational and
computational studies point to the fundamental role of some form of quasar
feedback for establishing them
\citep[e.g.][]{2001ApJ...554L.151B,2004ApJ...600..580G,2004MNRAS.347..144S,2005ApJ...620L..79S,2005MNRAS.363L..91C,2005MNRAS.358L..16K,2005Natur...433..604D,2006MNRAS.370..645B,2006MNRAS.370..289B,
2006MNRAS.365...11C, 2007MNRAS.382.1394M,
2007ApJ...665.1038C,2007MNRAS.380..877S, 2007ApJ...667..117T,
2007ApJ...669...45H, 2008MNRAS.385..161O}.

One fundamental aspect of the study of quasars is the form and evolution of
the Quasar Luminosity Function (QLF). Recent surveys, including SDSS
\citep{2000AJ....120.1579Y} and the 2dF QSO Redshift Survey
\citep{2002MNRAS.333..279L}, are now providing large samples over sufficient
redshift ranges that the QLF shape and evolution can be investigated in
detail. Also, numerous studies of the QLF have been made, covering the X-ray
\citep{ 1997MNRAS.291..324P, 2001A&A...369...49M, 2002ApJ...570..100L,
2003A&A...409...79F, 2003ApJ...598..886U, 2003ApJ...584L..57C,
2003ApJ...584L..61B, 2005AJ....129..578B, 2005ApJ...635..864L,
2008ApJ...679..118S, 2009A&A...493...55E, Yencho2009}, optical
\citep{2003A&A...408..499W, 2004MNRAS.349.1397C, 2006AJ....131.2766R}, radio
\citep{2005MNRAS.357.1267C, 2005A&A...434..133W}, and IR
\citep{2006A&A...451..443M, 2006ApJ...638...88B} bands.  Overall these studies
suggest that the spatial density of quasars undergoes a luminosity-dependent
evolution, with the density of more luminous quasars peaking at higher
redshift than the less luminous populations.

Theoretical investigation of the QLF has been done using semi-analytical
models \citep[e.g.][]{2000MNRAS.311..576K, Volonteri2003, 2003ApJ...595..614W,
2004ApJ...600..580G, 2007MNRAS.382.1394M, Marulli2008, Bonoli2009}.  Since these
models do not self-consistently follow black hole growth, the AGN lightcurves
and luminosity have to be calculated via a specified prescription. The
predominant method for modeling the quasars in this context is to treat
quasars as radiating at a fixed fraction of their Eddington luminosity for a
characteristic time-scale after a galaxy merger before shutting off completely
due to feedback effects. The determination of the characteristic time-scale
varies between methods. For example, \citet{Haiman2004} assume quasar
radiation at the Eddington luminosity for a fixed time-scale of $2 \times 10^7$
years for the radio-loud lifetime of the quasars.  \citet{Volonteri2003}
assume the quasar will maintain Eddington accretion until it has accreted a
total mass proportional to the fifth power of the circular speed of the merged
system.  \citet{2003ApJ...595..614W} adopt a model where the quasars radiate at a fixed
fraction of the Eddington luminosity for the dynamical time of the galactic
disc, at which point the gas has been given more energy than its binding
energy.  These methods have produced promising
results, but are all based on variants of the simple on-off model.

\citet{2005ApJ...630..705H, 2005ApJ...630..716H, 2005ApJ...625L..71H,
  2006ApJS..163....1H, 2006ApJ...639..700H} took a different approach to
  modeling the QLF by analyzing the light curves of quasars in hydrodynamical
  galaxy merger simulations which included black hole growth, accretion and
  feedback \citep[see][]{2005Natur...433..604D, 2005MNRAS.361..776S}, and used
  the results to express the quasar lifetime as a differential time a quasar
  spends radiating in a logarithmic luminosity bin \citep{2005ApJ...630..705H,
  2005ApJ...630..716H, 2005ApJ...625L..71H}.  The quasar lifetimes were fit to
  a Schechter function dependent on both peak and current luminosity.  In this
  way, quasars were modeled using detailed predictions from hydrodynamic
  simulations for their lightcurves and were shown to radiate at a range of
  luminosities both at and below their peak, rather than being restricted to
  radiating at a primarily constant peak luminosity.  

   Hopkins et al.'s approach found that using the predicted form for
  quasar lifetime, the faint-end of the QLF could be explained by quasars
  radiating well below their peak luminosities, rather than by quasars with
  low peak luminosities.  In this case, to match the observational form of the
  QLF, the quasar creation rate must peak at the critical break luminosity of
  the QLF, with a very rapid drop-off for luminosities below the break
  \citep{2005ApJ...630..716H}.  This work provided a fundamentally different
  explanation for the physical source of the faint-end slope and the break
  luminosity while still reproducing the form and evolution of the observed
  QLF \citep{2006ApJ...639..700H}.  However, the conclusions were based upon
  data extracted from individual galaxy merger simulations and have yet to
  be investigated with cosmological hydrodynamical simulations.

In this paper we analyse fully cosmological hydrodynamic simulations which
directly include modeling of black hole growth, accretion and associated
feedback processes (as well as the dynamics of dark matter, dissipation, star
formation and stellar feedback) and make predictions for the quasar luminosity
function and its evolution.  The simulations are currently among the largest,
highest-resolution hydrodynamical simulations which include gas hydrodynamics,
and have been shown already to reproduce many aspects of the black hole
evolution, such as the mass function and accretion rate distribution, and in
particular the assembly and evolution of the black hole galaxy correlations
\citep{2008ApJ...676...33D}. In this paper we compare the black hole
luminosity functions and their evolution from the simulations with appropriate
observations in various energy bands. This is both an important test for
assessing the value of the simulations and for providing a physical context
within which to interpret the observations and quasar evolution in general.

In Section 2 we describe the numerical modeling for the black holes accretion and
luminosity (Section 2.1) and the simulation parameters used (Section 2.2).  In Section 3 we
present the results for the black hole luminosity function, comoving number
density evolution, and luminosity density evolution, and compare with
observational data. In Section 4 we discuss the implications of our simulation on
the hard X-ray background, and in Section 5 we summarize and discuss our results.

\section{Method}

\subsection{Numerical simulation}

In this study, we analyse the set of simulations published in
 \citet{2008ApJ...676...33D}. Here we present a brief summary of the simulation
 code and the method used. We refer the reader to \citet{2008ApJ...676...33D}
 for all details.

The code we use is the massively parallel cosmological TreePM--SPH code
{\small Gadget2} (Springel 2005), with the addition of a multi--phase
modelling of the ISM, which allows treatment of star formation (Springel \&
Hernquist 2003), and black hole accretion and associated feedback processes
(Springel et al. 2005, Di Matteo et al. 2005).

Black holes are simulated with collisionless particles that are created in
newly emerging and resolved groups/galaxies.  A friends--of--friends group
finder is called at regular intervals on the fly (the time intervals are
equally spaced in log $a$, with $\Delta \log{a} = \log{1.25}$), and employed
to find groups of particles. Each group that does not already contain a black
hole is provided with one by turning its densest particle into a sink particle
with a seed black hole of fixed mass, $ M = 10^5 h^{-1}$\,M$_\odot$. The
black hole particle then grows in mass via accretion of surrounding gas
according to $\dot{M}_{\rm BH} = \frac {4 \pi G^2 M_{\rm BH}^2 \rho}{(c_s^2 +
v^2)^{3/2}}$ \citep{1939PCPS...35..405H, 1944MNRAS.104..273B,
1952MNRAS.112..195B}, and by merging with other black holes.

\begin{figure}
  \centering
  \includegraphics[width=9cm]{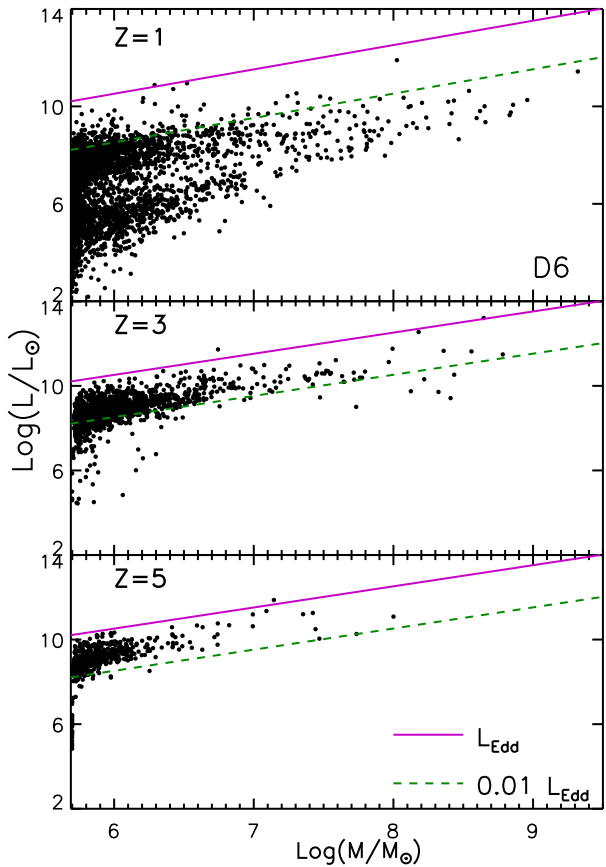}
    \caption{Relation between mass and bolometric luminosity for black holes in the D6 simulation for
    redshifts 1, 3, and 5.  The lines show $\rm{L_{Edd}}$ (solid pink)
    and $\rm{0.01 L_{Edd}}$ (dashed green).}
    \label{masslum}
\end{figure}

For the simulations used here it is assumed that accretion is limited
to a maximum of 3 times the Eddington rate. Note, very few sources accrete at
this critical value, as seen in Fig. \ref{masslum}.

	The accretion rate of each black hole is used to compute the
	bolometric luminosity, $L = \eta \dot{M}_{\rm BH} c^2$
	\citep{1973A&A....24..337S}.  Here $\eta$ is the radiative efficiency,
	and it is fixed at 0.1 throughout the simulation and this analysis.
	Some coupling between the liberated luminosity and the surrounding gas
	is expected: in the simulation 5 per cent of the luminosity is
	(isotropically) deposited as
	thermal energy in the local black hole kernel, acting as a form of feedback energy \citep{2005Natur...433..604D}.

	Note that to derive luminosities in specific wavebands (consistent with
	the observational constraints), we need to apply a bolometric
	correction to our quasar luminosities. We apply the bolometric
	correction from \citet{2007ApJ...654..731H} (consistent with
	\citep{2004MNRAS.351..169M}):

\begin{equation}
\frac{L}{L_{band}} = c_1 \left (\frac{L}{10^{10} L_\odot} \right )^{k_1}+c_2
\left (\frac{L}{10^{10} L_\odot} \right )^{k_2}
\label{bolcor}
\end{equation}

where c1 =(6.25, 17.87, 10.83), c2 = (9.00, 10.03, 6.08), k1 = (-0.37, 0.28,
0.28), k2 = (-0.012, -0.020, -0.020) for B-Band, 0.5-2 keV Soft X-ray band,
and 2-10 keV Hard X-ray band, respectively.

\subsection{Simulation parameters}
\begin{table}
\caption{Numerical Parameters}
\begin{tabular}{c c c c c c}

  \hline
  \hline
  
  Run & Boxsize & $N_p$ & $m_{\rm DM}$ & $m_{\rm gas}$ & $\epsilon$ \\
   & $h^{-1} {\rm Mpc}$ & & $h^{-1} M_{\odot}$ & $h^{-1} M_{\odot}$ & $h^{-1}
   {\rm kpc}$ \\
  
  \hline
  
  D4 & 33.75 & $2 \times 216^3$ & $2.75 \times 10^8$ & $4.24 \times 10^7$ & 6.25 \\
  D6 & 33.75 & $2 \times 486^3$ & $2.75 \times 10^7$ & $4.24 \times 10^6$ & 2.73 \\
  E6 & 50 & $2 \times 486^3$ & $7.85 \times 10^7$ & $1.21 \times 10^7$ & 4.12 \\

\hline

\multicolumn{6}{l}{$N_p$: Total number of particles} \\
\multicolumn{6}{l}{$m_{\rm DM}$: Mass of dark matter particles} \\
\multicolumn{6}{l}{$m_{\rm gas}$: Initial mass of gas particles} \\
\multicolumn{6}{l}{$\epsilon$: Comoving gravitational softening length} \\

\end{tabular}
\label{param}
\end{table}
Three simulation runs are analysed in this paper to allow testing for
resolution effects. The main parameters are listed in Table \ref{param}.  The
three runs were of moderate volume, with boxsizes of side length $33.75 h^{-1}
{\rm Mpc}$ (D6 and D4 simulations), and $50 h^{-1} {\rm Mpc}$ (E6). For the D6 and E6 runs
$N_p = $ $2 \times 486^3$ particles were used, and the D4 used $2 \times
216^3$. The moderate boxsizes prevent the simulations from being run below $z\sim
1$ ($z\sim 0.5$ for D4 run) to keep the fundamental mode linear, but provide a large enough scale to
produce sufficiently luminous sources, albeit rare. The limitation on
the boxsizes is necessary to allow for appropriate resolution to carry out the
subgrid physics in a converged regime (for further details on the simulation
methods, parameters and convergence studies see \citet{2008ApJ...676...33D}
and also the discussion at the end of this paper).

\begin{figure}
  \centering
  \includegraphics[width=8.5cm]{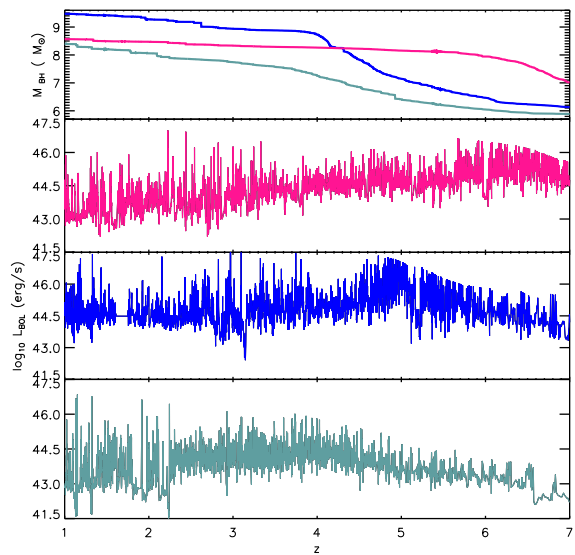}
    \caption{Example of lightcurves between z=1 and z=7 for three massive
      black holes in the D6 simulation. The top panel shows the growth of
      the black hole masses and bottom three panels (in corresponding colors)
      show the associated lightcurves for their bolometric luminosity.}
    \label{lightcurve}
\end{figure}

\section{Results}
\subsection{Mass-Luminosity Relation}
In order to illustrate the range of properties of the black hole population in
the simulations, in Fig. \ref{masslum} we show the relation between black
hole mass and luminosity for the whole sample of objects in the D6 simulation
(at $z= 1, 3, \: \rm{and} \: 5$).  Note the D4 and E6 mass-luminosity relations are
not plotted, but both produced similar results. There is some correlation,
albeit weak, between luminosity and mass of black holes, however in most
regimes a significant scatter is seen, implying a fair range of luminosities
for a fixed black hole mass. This is the direct result of our simulations and
in particular the complex lightcurves associated with the accretion history (and
the evolution of the gas supply) which is followed in detail for all the black
holes in our simulations. As an example, in Fig. \ref{lightcurve} we show the
black hole mass assembly history for three specific black holes in the D6
simulations and their associated lightcurves (in terms of bolometric
luminosity). The high level of variability in the lightcurves is induced by
the detailed hydrodynamics, interplay between gas inflows, associated
accretion and feedback processes self-consistently modelled in the simulations
(see also \citet{2008ApJ...676...33D} for more examples of accretion and merger
histories of black holes in the simulations).  This implies that in turn we
expect the same black holes, at different stages of activity, to contribute to
different regions of the luminosity function. 

\subsection{Luminosity Functions}
\begin{figure}
  \centering
  \includegraphics[width=8cm]{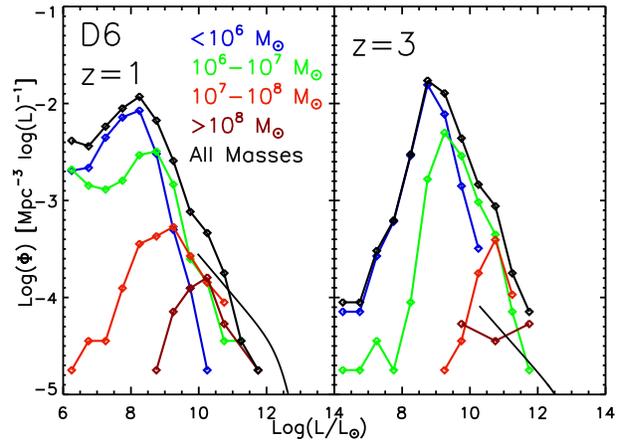}
    \caption{The luminosity function as computed for the D6 simulation at
    redshifts 1 and 3 for black holes restricted to the following ranges:
    Blue: $< 10^6 M_\odot$; Green: $10^6-10^7 M_\odot$; Red: $10^7-10^8
    M_\odot$; Brown: $> 10^8 M_\odot$; Black: Full mass range.  The \citet{2007ApJ...654..731H} best-fitting bolometric QLF has also been plotted for comparison (Solid black line).}
    \label{LumFuncMassBin}
\end{figure}

To illustrate the effect of different black hole populations to the QLF in
detail, in Figure  \ref{LumFuncMassBin} we plot the relative contribution from
the different black hole mass ranges to the luminosity function at $z= 1$ and
3. At $z = 1$, the black holes with mass below $10^7 M_\odot$ provide the
dominant contribution to the luminosity function for luminosities below
$10^{9.5} L_\odot (10^{43} \rm{erg} \: \rm{ s}^{-1})$, while higher masses
dominate at larger luminosities. At higher redshift, the low mass black holes
are the dominant contribution up to a higher luminosity ($10^{10.5} L_\odot$
at $z=3$), and have a more significant contribution to higher luminosities
than they do at low redshift. Note that black holes with masses $ < 10^7
M_{\odot}$ give rise to a significant steepening of the luminosity function
below $\sim 10^{10} L_{\odot}$ (although this is sometimes below the current
observational limits, it is an important effect in our results). In our
numerical simulations we do not expect to resolve the accretion history on to
the lowest mass black holes (as also shown by increased scatter in
Fig. \ref{masslum}), where the gas dynamics are well resolved only well beyond
the black hole accretion radius. For this reason, as well as the fact that the
low-mass black holes correspond primarily to recently inserted seed particles
which have yet to undergo critical accretion phases (i.e. dependent on our
initial choice of this parameter), we will use only black holes with $M_{\rm
  BH} > 10^6 M_{\odot}$ for the rest of our analysis.

\begin{figure*}
  \centering
  \includegraphics[width=16cm]{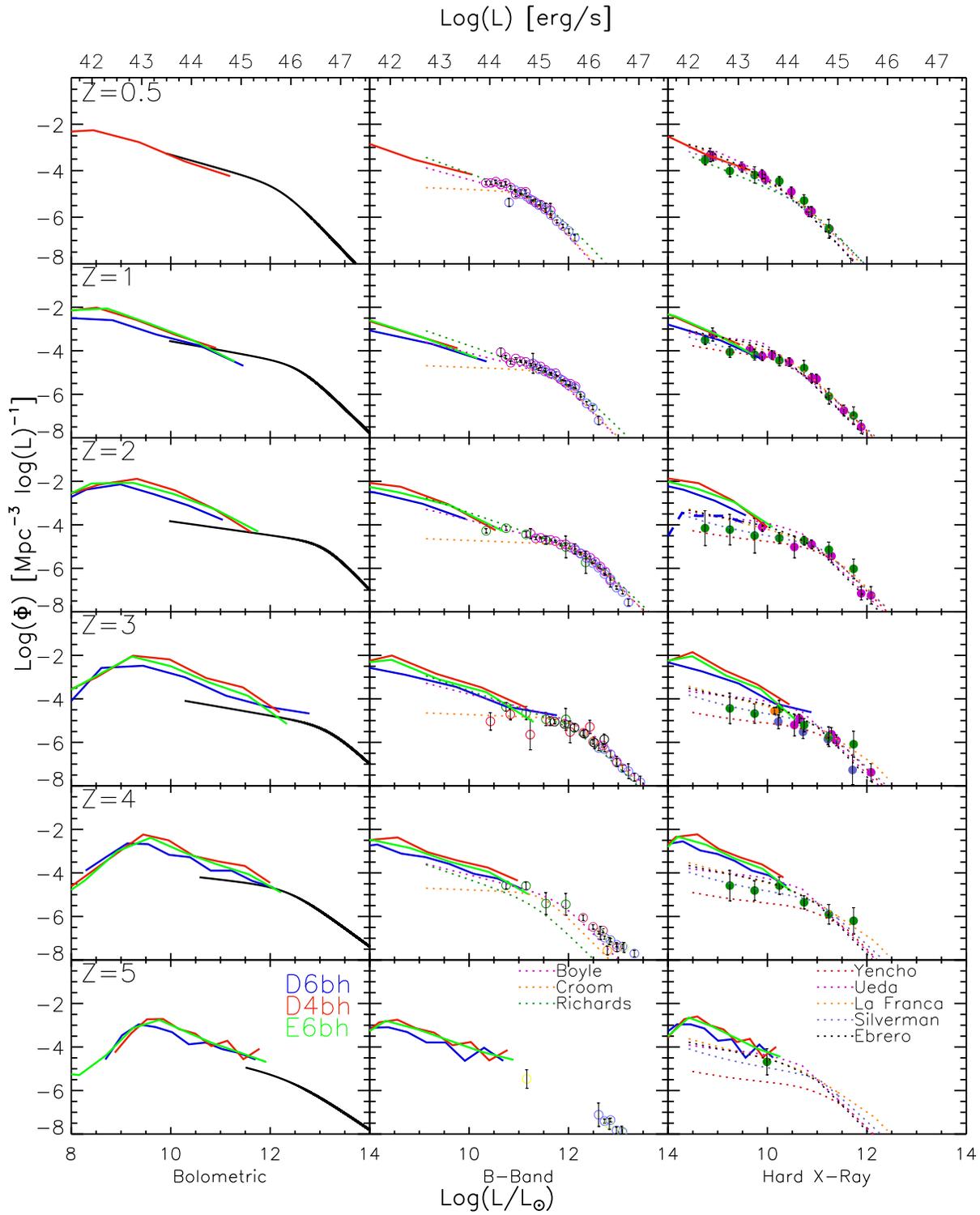}
    \caption{The black hole luminosity function for all three
    simulations (blue - D6 simulation; red - D4 simulation; green - E6
    simulation) using sources with $M_{\rm BH} > 10^6 M_{\odot}$.
    The first column is the bolometric QLF computed directly from the
    simulation.  The solid black line is the double-power law QLF
    function given in \citet{2007ApJ...654..731H}.  The second and third
    columns show the luminosity function after applying a bolometric
    correction (Eq. \ref{bolcor}) to produce the B-Band and Hard X-ray band.  Open
    circles for optical bands are datapoints 
    from the following studies: Bright purple - \citet{2005MNRAS.360..839R};
    Blue - \citet{2006AJ....131.2766R}; Dark Green -
    \citet{2003A&A...408..499W}; Red - \citet{2004ApJ...605..625H}; Yellow - \citet{2004ApJ...600L.119C};
    Orange - \citet{1995AJ....110.2553K}; Dark Purple - \citet{1995AJ....110...68S}; Bright
    Green - \citet{2001AJ....122.2833F, 2001AJ....121...54F,
    2003AJ....125.1649F, 2004AJ....128..515F};
    Black - \citet{2006astro.ph..4373S}. Closed circles for hard X-ray bands are
    datapoints from the following studies: Pink -
    \citet{2003ApJ...598..886U}; Blue - \citet{2005ApJ...618..123S};
    Green - \citet{2005AJ....129..578B, 2003AJ....126..632B,
    2003ApJ...584L..61B}; Orange - \citet{2005MNRAS.360L..39N}.  Dotted lines
    in the hard X-ray column are best-fitting LDDE functions from the following
    studies: Pink - \citet{2003ApJ...598..886U}; Orange -
    \citet{2005ApJ...635..864L}; Purple - \citet{2008ApJ...679..118S}; Black -
    \citet{2009A&A...493...55E}; Red - \citet{Yencho2009}.  Dotted
    lines in the B-Band column are best-fitting LDDE functions from the following
    studies: Pink - \citet{2000MNRAS.317.1014B}; Orange -
    \citet{2004MNRAS.349.1397C}; Green - \citet{2005MNRAS.360..839R}. The
    dashed line in the hard X-ray at z=2 is the D6 QLF if only
    $M_{\rm BH}>10^{7.5} M_{\odot}$ are included.}
    \label{LuminosityFunc}
\end{figure*}

Fig. \ref{LuminosityFunc} shows the predictions from our simulations for the
AGN luminosity functions for $z=0.5, 1, 2, 3, 4, 5$, (note that only the D4
simulation is used at $z=0.5$). The first column shows
the bolometric luminosity function derived directly from the simulations. The
second and third columns show the luminosity function after applying the
bolometric correction (Eq.1) to obtain the B-band (second column) and 2-10
keV hard X-ray band (third column) QLFs.  Along with our predictions, we plot
the observational data and the best-fitting QLFs from several studies for the hard
X-ray and optical bands (see Fig. \ref{LuminosityFunc} caption for complete
list of references)\footnote{Note that both \citet{2009A&A...493...55E} and
\citet{Yencho2009} considered 2-8 keV rather than 2-10 keV, so their
functions were adjusted using a photon index of $\Gamma = 1.8$ to maintain a
consistent definition of the hard X-ray band. In addition, neither
\citet{2009A&A...493...55E} nor \citet{Yencho2009} consider
absorption, whereas \citet{2003ApJ...598..886U}, \citet{2005ApJ...635..864L},
and \citet{2008ApJ...679..118S} all use absorption corrected data.}.

The bolometric luminosities are compared to the best-fitting function computed by
\citet{2007ApJ...654..731H} who compiled all available data from observational
studies across several bands, including the optical, mid-infrared, hard and
soft X-ray bands, and fitted them to a double power law function (see
\cite{2007ApJ...654..731H} for the function and the table of redshift-dependent
parameters). The best-fitting function is plotted consistent with the range of observed bolometric luminosities. This is why the minimum luminosity shown in these fit
functions is redshift-dependent, and ranges from $9.97 L_\odot$ at z=1 to
$11.53 L_\odot$ at z=6.

Comparing observed and predicted LFs, it is apparent that the simulations can
only reproduce the 'faint-end' of the LF: this is expected as the number
density of AGNs in the 'bright-end' is simply too low to be accessible in our
simulated volumes. Thus our predictions are limited to a relatively small
range of luminosities which can be compared directly to observational data,
and the largest overlap is in the X-ray band, rather than in the B-band due to
the significantly fainter AGN populations in the former. Related to this is
the lack of predictions for the knee of the QLF, which
occurs at a higher luminosity than the simulations produce.

Within the range of comparison, our predictions agree well with the data. Our
simulations are fully consistent with the constraints from the B-band (albeit
with very limited region of overlap). In the hard X-ray band, at $L \sim
10^{10} \L_{\odot}$ ($L \sim 10^{43.5}$ erg s$^{-1}$), close to the maximum
luminosities probed with our simulations, there is also very good
agreement. For $z \le 1$ the overall shape and slope of the faint-end is also
reproduced remarkably well. At $ 2 \le z \le 4$ however, the slope predicted from
the simulation is typically steeper than in the observed LFs.  At $ z >4$, if
compared to the fits of the observations, the slopes are again consistent. The
same result is found in the comparison with the bolometric luminosity
functions (where indeed the hard X-ray data significantly dominates Hopkins'
fits in the low luminosity end) and where the discrepancy in the slope is the
greatest at $z \sim 2$.

It is promising that the simulations agree well with the data at $z \le 1$
where the observations in the 'faint-end' are most complete and the bolometric
corrections (which are derived from the local samples and have no redshift
dependence) are most appropriately applied. The larger number of AGNs at the
faintest luminosities (hence the steeper slopes) at higher redshifts may have
two possible implications. One is that there is still a population of the
faint, possibly heavily obscured AGN above $z=2$ that have not yet been
detected. Alternatively, our simulations are actually overproducing the faint
AGN population due to lack of appropriate resolution or appropriate feedback
physics (either due to stellar processes or AGN) in the faint, low mass black
hole population. The latter possibility is also hinted at from our results in
Fig. \ref{LumFuncMassBin} and the fact that the D6 predictions (highest
resolution simulation) typically predict the flattest slopes (albeit barely as
the convergence between the predictions for the three simulations is good). We
will investigate this further in the course of the paper.

\begin{figure*}
  \centering
  \subfigure
  {
    \label{ModelVariations:a}
    \includegraphics[width=8.0cm]{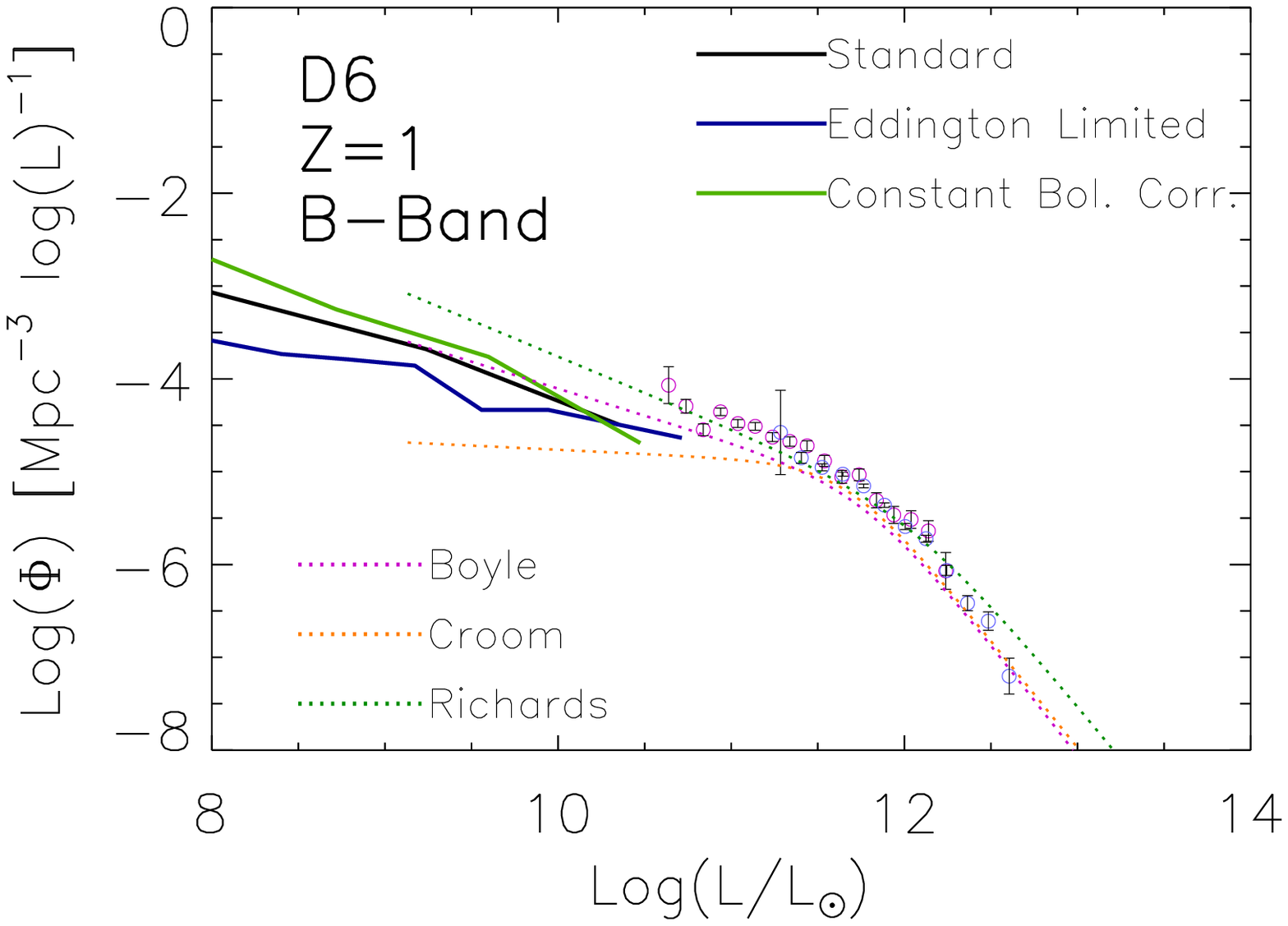}
  }
  \hspace{1cm}
  \subfigure
  {
    \label{ModelVariations:b}
    \includegraphics[width=8.0cm]{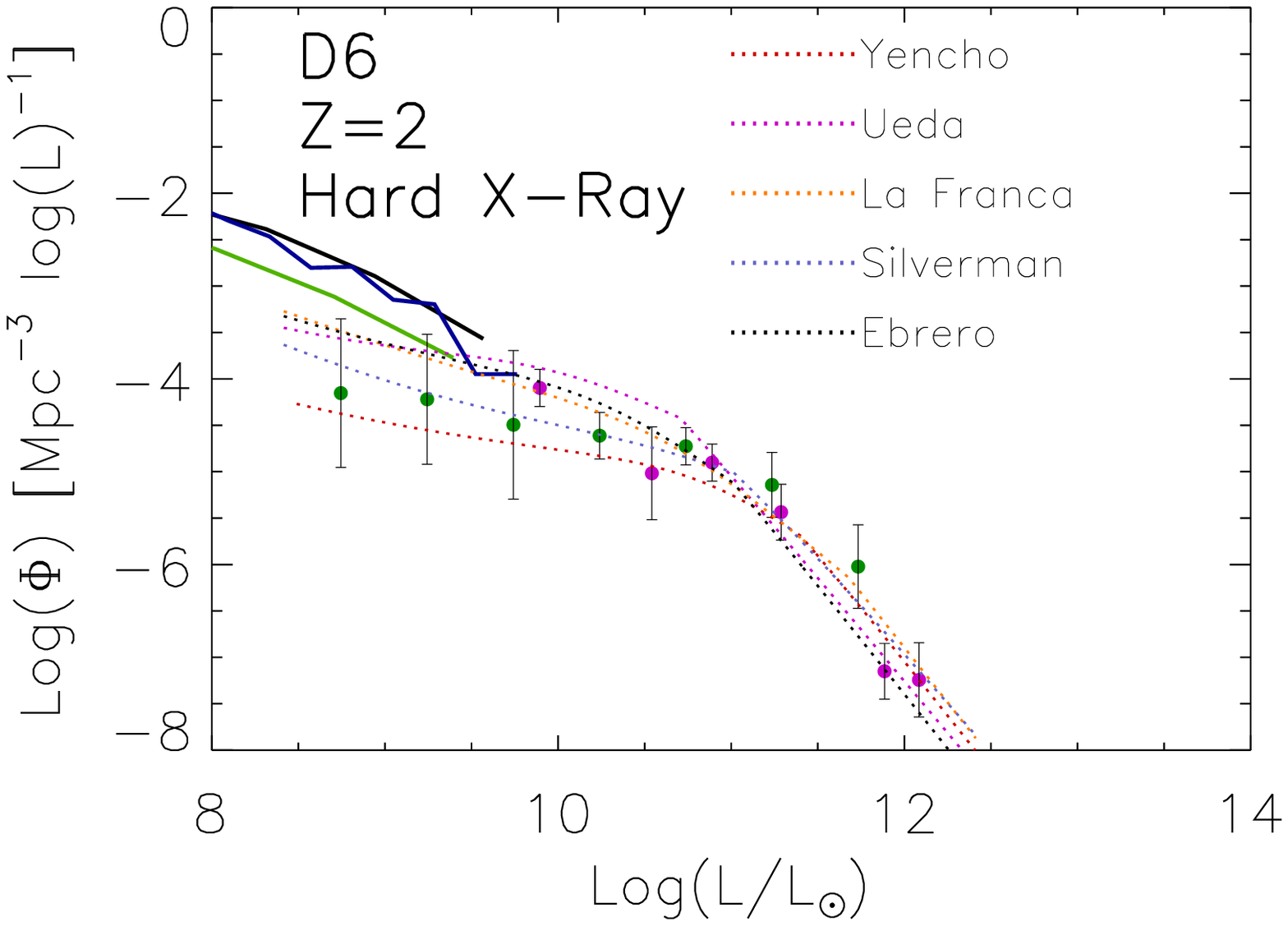}
  }
  \caption{The QLF for D6 simulation for B-Band at z=1 (left) and Hard X-ray
  at z=2 (right) using different model parameters: Black - same parameters as
  Fig. \ref{LuminosityFunc}; Blue - sources with $\rm{L} > \rm{0.01 L_{Edd}}$;
  Green - luminosity-independent bolometric correction.  See
  Fig. \ref{LuminosityFunc} for references of observational datapoints and
  best-fitting functions.}
    \label{ModelVariations}
\end{figure*}

\subsubsection{Model dependent effects on the predictions for QLF}

 Even though our simulations predict directly an accretion rate for
  each of the BHs (from the hydrodynamics), at any given time there are two
  major model dependent assumptions that we have made in order to the
  translate the accretion rate into a luminosity in a particular waveband: one
  is the assumption of a fixed radiative efficiency of ten per cent, and the
  second is the use of an empirically derived bolometric correction.  Here we
  wish to investigate the main effects on the QLF predictions of varying these
  two assumptions.  In Figure \ref{ModelVariations} we demonstrate these
  effect by showing the B-Band QLF at z=1 (left) and the hard X-ray band QLF
  at z=2 (right). With these two panels we are able to fully illustrate the
  relevance of the effects.

  When calculating black hole luminosities (as described in Section 2.1), we
  are assuming that all sources have equal radiative efficiency and set it to
  0.1. However, it has been suggested (even though the precise changes in
  accretion physics at low accretion rate remain somewhat uncertain) that
  sources accreting at a sufficiently low Eddington fraction (typically at
  $\approx 0.01$ Eddington) are expected to transition to a radiatively
  inefficient state with associated changes in the spectral energy distribution
  (SED) that are dramatically different.  These radiatively inefficient
  accretion models \citep[e.g.][and references therein]{Narayan2005, QuataertNarayan1999,
    YuanNarayan2004} (and also
  observations of both AGN and X-ray transients) indicate that such transitions
  occur around accretion rates of 1\% the Eddington value.  The simplest way
  to investigate the overall effect this may have on our QLF predictions is to
  eliminate all sources accreting below $0.01 \rm{L_{Edd}}$ (blue line).  As
  shown in Figure \ref{masslum}, most sources are above this cut-off
  luminosity at $z > 1$, so we expect a minimal effect on the QLF above this
  redshift.  Eliminating low luminosity sources for $z\le 1$ leads to a flattening
  of the QLF slope (Fig. \ref{ModelVariations}), as most sources are actually
  below this threshold at this point (as indeed generally expected that such
  modes of accretion will be well below the quasar peak).  In short, low radiative
  efficiency accretion is expected to lead to some flattening of the QLF
  at $z \le 1$. This effect therefore would not help flattening the QLF
  function at higher redshift and therefore better reconcile our predictions
  with observations.
\footnote{
Additionally, because our simulations use a single feedback model for
  all black holes, they do not model separate 'quasar' and 'radio' modes.  In addition to having an effect on the radiative efficiency
  used to determine the BH luminosity, the inclusion of a radio mode will have
  a quenching effect during the simulation, due to the radio mode suppressing
  inflow of cooling gas \citep{2006MNRAS.365...11C}.  As the majority of our
  sources at $z>1$ are accreting above 0.01 times the Eddington accretion rate, it is
  only at low redshifts (at or below $z \approx 1$, which is the limit of our
  simulations) that we would expect the radio mode to have a significant
  effect on black hole growth.  Additionally, \citet{2007MNRAS.380..877S}
  found that, although the effect of the radio mode does become large at $z <
  1$, the bulk of black hole growth is always during the quasar mode (with the
  quasar mode accretion contributing 95\% of the integrated black hole mass
  density), and that modeling a separate radio mode has negligible effect on
  $M_{\rm{BH}}-M_\star$ and $M_{\rm{BH}}-\sigma_\star$ relations. }
  
Our predictions for the various wavebands are of course dependent on
  the form of the bolometric correction used (the one adopted here is shown in
  Eq. 1). Even though there is a luminosity dependence in our bolometric
  correction, it is less well-constrained at low luminosities (also for the
  reasons discussed above), where the majority of our sources lie.  To explore
  the effects that the luminosity dependence has on our results in Figure \ref{ModelVariations} (green line) we show the QLFs for a
  luminosity-independent bolometric correction, using the value of the
  correction factor evaluated at $L=10^{12} L_\odot$, where the correction
  factor is best constrained.  Doing this has a small effect on the B-Band
  QLF, where in any case (see Eq.1) the correction has a small dependence on
  luminosity. In the hard X-ray band, however, a luminosity
  independent correction produces significantly lower magnitude for the QLF, which
  more closely matches the observational data.  This illustrates that the
  exact form of the bolometric correction, particularly the form of its
  luminosity dependence, may have a strong effect on our final results.

\subsection{Comoving Number Density Evolution}

\begin{figure*}
  \centering
  \subfigure
  {
    \label{Den:a}
    \includegraphics[width=8.0cm]{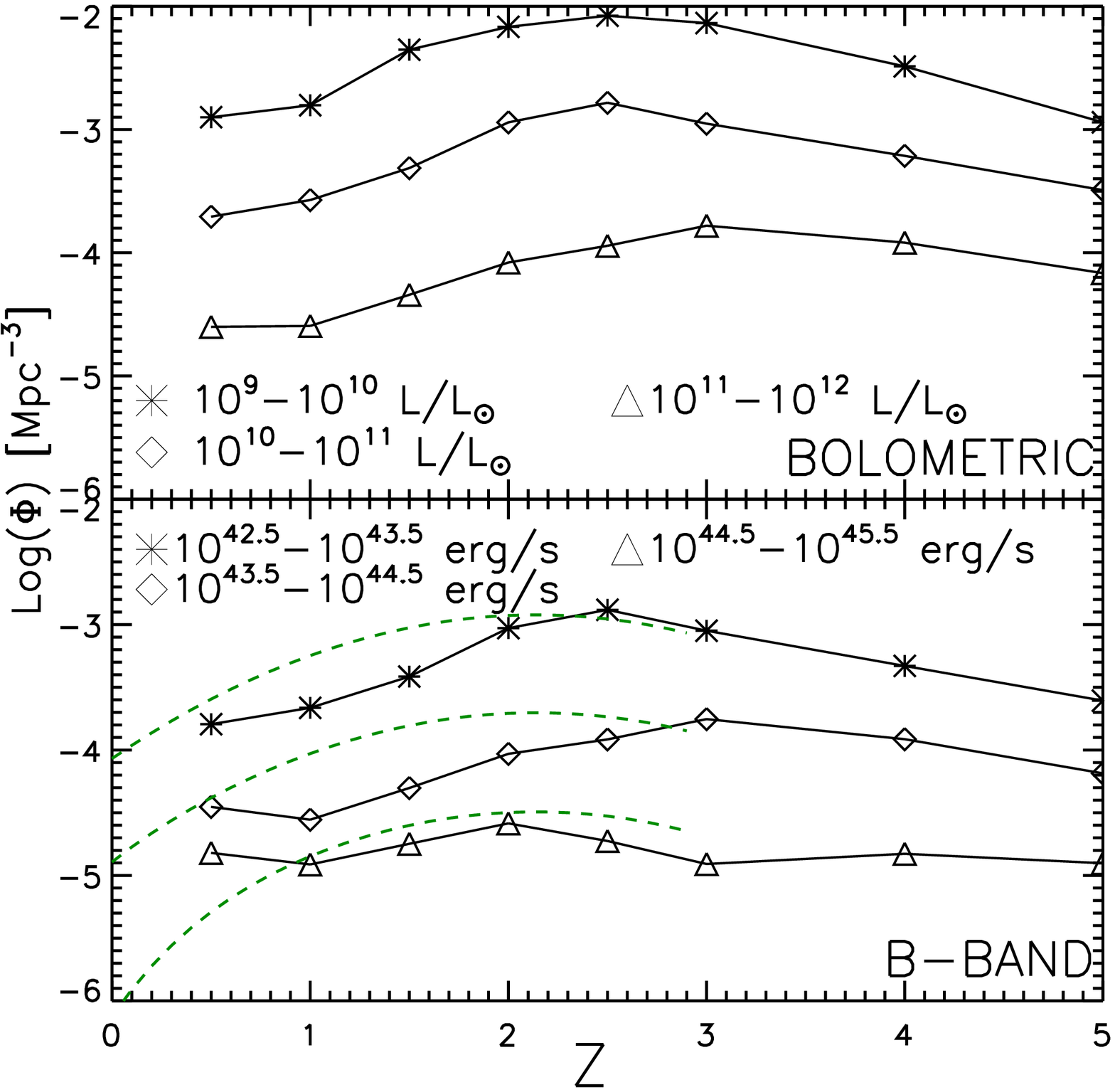}
  }
  \hspace{1cm}
  \subfigure
  {
    \label{Den:b}
    \includegraphics[width=8.0cm]{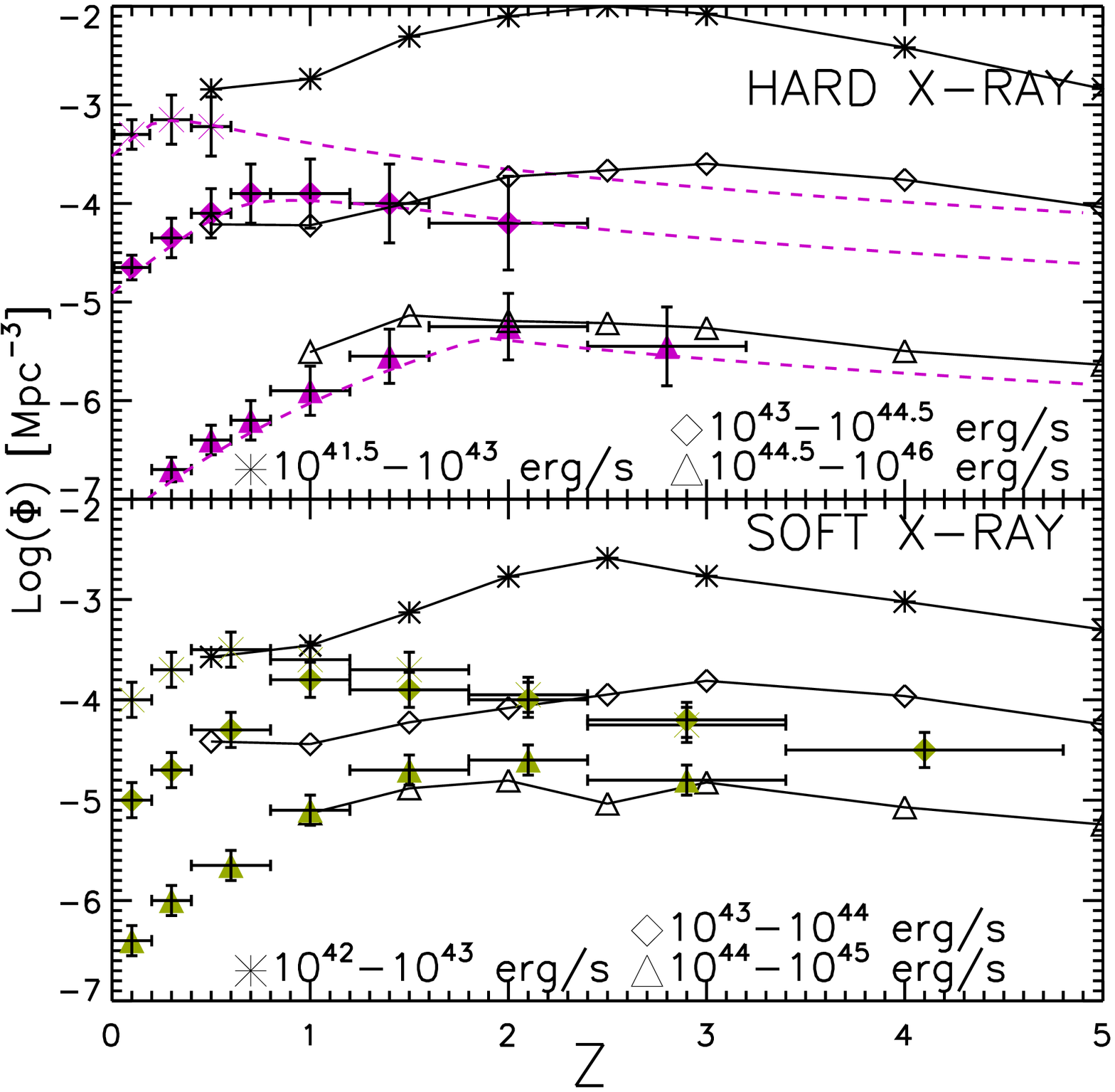}
  }

    \caption{The comoving number density evolution of quasars for $M_{\rm BH} > 10^6  M_{\odot}$
      averaged across the three simulations (except at $z=0.5$, where only the
      D4 simulation was available).  A linear extrapolation has been
      applied to the simulation to allow our luminosity function to be
      integrated over the given luminosity ranges, and be compared directly to
      observational data from \citet{2003ApJ...598..886U} (hard X-ray - purple
      triangles)
      and \citet{2005A&A...441..417H} (soft X-ray - green triangles).  We have
      also plotted the predictions from the best-fitting functions from
      \citet{2005MNRAS.360..839R} (optical - green dashed line) and
      \citet{2003ApJ...598..886U} (hard X-ray - purple dashed line).  Note
      the optical observational data were limited to redshifts below $z=3$, so
      we have only plotted those curves for $z < 3$.  Additionally, the soft
      X-ray data have been adjusted for obscuration as in \citet{2006ApJ...639..700H}.}
    \label{Den}
\end{figure*}

      	The quasar comoving number density evolution is plotted in Fig.
	\ref{Den}. This is derived by integrating the luminosity functions
	plotted in Fig. \ref{LuminosityFunc} (we average over all three
	simulations). Again we plot the predictions for the bolometric,
	B-band, hard X-ray and in this case soft X-ray band also. For the
	latter two bands we also show the observational constraints from
	\citet{2003ApJ...598..886U} (hard X-ray) and
	\citet{2005A&A...441..417H} (soft X-ray). Note that, following
	\citet{2006ApJ...639..700H}, the normalization of the soft x-ray data
	has been multiplied by 10 to adjust for obscuration in this band (this
	adjustment is somewhat model dependent, but provides a first
	approximation for comparison).  We have also plotted the predictions
	from the best-fitting functions of \citet{2005MNRAS.360..839R} (B-Band)
	and \citet{2003ApJ...598..886U} (Hard X-ray).  The B-Band function is
	terminated at $z = 3$ since the fits were based only on sources below
	this redshift.  In some cases a linear extrapolation (to
	higher luminosities) was
	applied to the simulation to allow the range of
	integration to match the observational data (given in specific
	luminosity bins).

	In virtually every band, the quasar number density from the
	simulations peaks at $z \sim 2.5$ and as expected, their number
	density is dominated by the lower luminosity populations. When
	comparing to the X-ray data (both hard and soft bands), we again find
	there is good agreement with the observed evolution in the
	intermediate/high luminosity ranges. Consistent with the results from
	the luminosity functions, the evolution in the lowest luminosity range
	implies larger number densities in the soft X-ray band above $z \ge
	2$.  The hard X-ray data from \citet{2003ApJ...598..886U} for the
	lowest luminosity range is limited to $z < 0.5$, preventing direct
	comparison, however the best-fitting function's extrapolation shows that
	we may have a similar overestimate for the hard X-ray band.

\begin{figure}
  \centering
    \includegraphics[width=8.0cm]{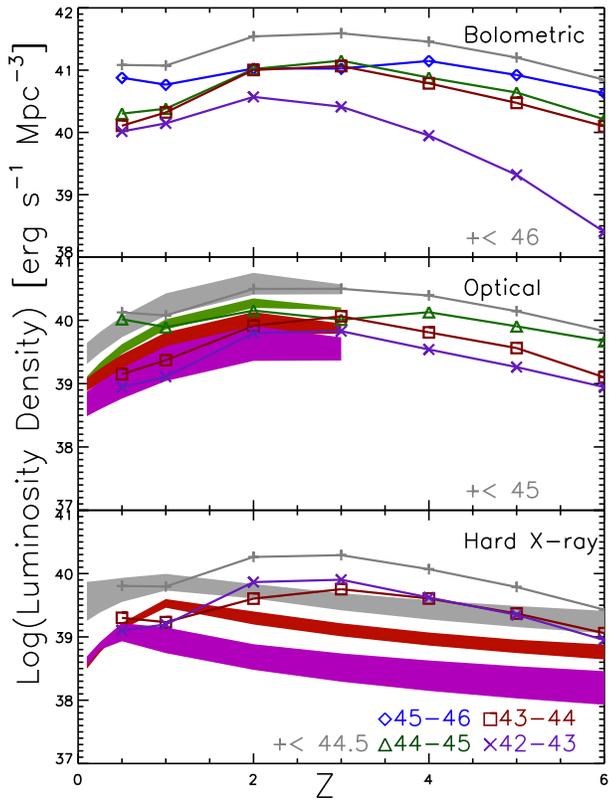} 
    \caption{The total luminosity density evolution averaged across the three
    simulations (except at $z=0.5$ where only the D4 simulation was used).
    Top panel - bolometric prediction computed directly from simulation; remaining
    plots - optical and hard X-ray predictions obtained by applying bolometric
    corrections to the bolometric luminosities.  Grey crosses show the luminosity density
    evolution below a certain cutoff luminosity (chosen at approximately the
    lowest break luminosity for the given band to restrict our predictions to the
    faint-end).  The remaining (colored) symbols show the emissivity produced
    by black holes within the following luminosity ranges: Blue diamonds - $ 10^{45} \: \rm{erg} \: s^{-1}
    < L < 10^{46} \: \rm{erg} \: s^{-1}$; Green triangles - $ 10^{44} \: \rm{erg} \:
    s^{-1} < L < 10^{45} \: \rm{erg} \: s^{-1}$; Red squares - $ 10^{43} \: \rm{erg} \:
    s^{-1} < L < 10^{44} \: \rm{erg} \: s^{-1}$; Purple X - $ 10^{42} \: \rm{erg} \:
    s^{-1} < L < 10^{43} \: \rm{erg} \: s^{-1}$.  The shaded regions are the areas
    bounded by the best-fitting LDDE functions from \citet{2003ApJ...598..886U},
    \citet{2005ApJ...635..864L} and \citet{2009A&A...493...55E} for hard X-ray
    and for optical, \citet{2005MNRAS.360..839R} and
    \citet{2000MNRAS.317.1014B}.  Note the observational
    estimates for the optical band are only plotted for $z < 3$ since the
    observational data was limited to low redshifts.}
    \label{Emis}
\end{figure}

\subsection{Luminosity Density Evolution}
	
	In Fig. \ref{Emis}, we show the total luminosity density evolution
	from the simulated quasars with the appropriate observational
	constraints. A linear extrapolation of the QLF was made for the
	highest luminosity bins such that consistent luminosity ranges could
	be used across all simulations and redshifts. To cover the full range
	of observational constraints, we have used the best-fitting
	luminosity-dependent density evolution (LDDE) functions from the
	following studies to bound the shaded regions: in optical -
	\citet{2005MNRAS.360..839R} and \citet{2000MNRAS.317.1014B}; in hard
	X-ray - \citet{2003ApJ...598..886U}, \citet{2005ApJ...635..864L} and
	\citet{2009A&A...493...55E}. For these reasons, Fig. \ref{Emis} is
	less of a direct comparison between simulations (where we extrapolate
	somewhat) and observations (where we used model dependent fits to data
	as constraints).

	While the predicted peak in the total luminosity density is at $z\sim
	2.5$ across the various bands, there are some fairly marked differences
	in the objects that dominate the contribution to the total luminosity
	evolution. The bolometric luminosity density (top panel) is dominated
	by the highest luminosity bins, with the brightest objects peaking at
	$z \sim 4$ while the lower luminosity bins peak at $ z \sim 2$.  In the
	optical band (second panel), the luminosity density is still dominated
	by the brightest objects, comparable to the observational constraints.

	The most significant difference is in the X-ray band, where the low
	luminosity population produces most of the contribution to the
	luminosity density. Additionally, the peak in observed luminosity
	density is close to $z\sim 1$ rather than $z\sim 2.5$ as implied by
	the simulations.  The overall reversal of trends in the relative
	contributions in various bands is of course caused by the form of the
	bolometric correction used in Eq. \ref{bolcor}.

\begin{figure}
  \centering
    \includegraphics[width=8.0cm]{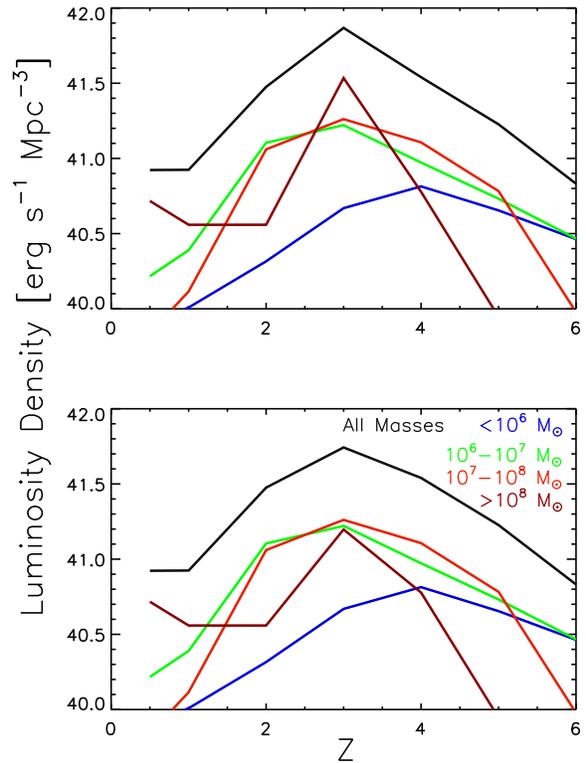}
    \caption{The total bolometric luminosity density evolution averaged across the three
    simulations (except at $z=0.5$ where only the D4 simulation was run), broken into mass bins. The upper plot includes all
    black holes, while the lower plot has neglected the outlying black
    hole in the D6 simulation at redshift 3.}
    \label{LumSumMassDep}
\end{figure}

	Finally, Fig. \ref{LumSumMassDep} shows the contribution to the
	bolometric luminosity density evolution as a function of black hole
	mass. The lower plot in Fig. \ref{LumSumMassDep} is identical to
	the upper plot except the single most luminous black hole from
	the D6 simulation at redshift 3 has been neglected (to explicitly show
	effects due to small statistics). We find that it is typically the
	midrange black holes masses ($10^6 M_{\odot} < M_{\rm BH} < 10^8
	M_{\odot}$) which provide the largest contribution to the luminosity
	density.  As one might expect, at higher redshifts, the low mass black
	holes provide a more significant contribution, which was expected due
	to the lack of high mass black holes (see also Fig. \ref{masslum}).

\section{Discussion}
One of our primary results from the luminosity function and the global number
density and luminosity density evolution is that our simulations are in good
agreement with the observational constraints but imply a larger number of low
luminosity X-ray sources (at $z > 2$) than observed.  In order to assess the
viability of our results we need to check whether this population may violate
the current constraints on the total X-ray background (XRB). The XRB intensity
is calculated according to \citep{1999coph.book.....P}

\begin{equation}
I_\nu = \frac{c}{4 \pi H_0} \int^{z_2}_{z_1} \epsilon_\nu ([1 + z] \nu_0 , z) \frac{dz}{(1
  + z)^2 \sqrt{(1 + \Omega_0 z)}}
\end{equation}

where the emissivity, $\epsilon_\nu$ is the hard X-ray emissivity shown in
  Fig. \ref{Emis}, $z$ is the redshift, [$z_1,z_2$] is the range of
  redshifts being considered, $\nu_0$ is the frequency at redshift zero, $H_0$
  is the Hubble Parameter at redshift zero, and $\Omega_0$ is the total
  density parameter at redshift zero.  A photon index of $\Gamma = 1.8$ was
  assumed when computing $\epsilon_\nu ([1+z] \nu_0)$ to account for the form
  of the power spectrum of the black holes.  Since we only have simulation
  information at discrete redshifts, we interpolate $\epsilon_\nu$ linearly between
  datapoints to compute the integral.

  We find that the total contribution to the 2-10 keV X-ray background from
  our simulated black holes is $I_{2-10 {\rm keV, D6}} =1.28 \times 10^{-11} \rm{erg \:
  s^{-1} cm^{-2} deg^{-2}}$ for the D6 simulation (recall the simulations
  are restricted to $z \ge 1$). This is well within the observed 2-10 keV hard X-ray
  background intensity, $ I_{2-10 {\rm keV, obs}} = 2.02 \pm 0.11 \times 10^{-11} \rm{erg
  \: s^{-1} cm^{-2} deg^{-2}}$ \citep{Moretti2003}. If we apply a linear
  extrapolation to the simulation to include an approximate contribution
  from $z < 1$, we predict a total XRB intensity of $I_{2-10 {\rm keV}} =1.8
  \times 10^{-11} \rm{erg \: s^{-1} cm^{-2} deg^{-2}}$, still below the observed
  value.

  When we do a similar calculation using the E6 and D4 simulations (which have
  lower resolution) we produce XRB intensities of $I_{2-10 {\rm keV}} =1.94
  \times 10^{-11}$ and $2.87 \times 10^{-11} \rm{erg \: s^{-1} cm^{-2}
  deg^{-2}}$  for the E6 and D4 simultions at $z > 1$, respectively.
  With the D4 simulation we are starting to violate the observed XRB even
  without including $z < 1$. This result is quite fundamental for our
  analysis as it indicates that the simulation resolution plays an important
  role in estimating the effect from the faintest sources (those that were
  found to be in excess of the observed LF) and that their contribution
  decreases with higher resolution (note in Fig. \ref{LuminosityFunc} the steepest slopes of LF
  are always predicted from the D4 run).  
  
\begin{figure}
    \centering
    \includegraphics[width=8.0cm]{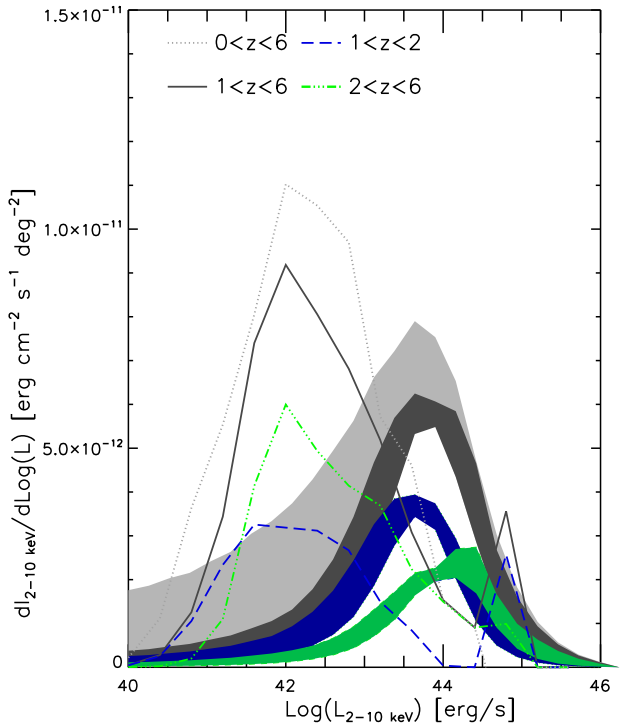}
    \caption{The differential contribution to the 2-10 keV hard X-ray background
    intensity as a function of luminosity, plotted for several redshift bins.  The
    shaded areas show the region of contributions bounded by the predictions by
    \citet{2003ApJ...598..886U}, \citet{2005ApJ...635..864L} and
    \citet{2009A&A...493...55E}.  The dotted line is an extrapolation of the
    simulation obtained
    by extending the luminosity density evolution from $z=1$ to $z=0$.}
    \label{BackgroundContributionVsLum}
\end{figure}

  Our simulations also predict that the QLFs extend to fainter fluxes than
  currently observed. We therefore compare our predictions for the unresolved
  fraction of the 2-8 keV XRB from the D6 simulation to test whether current
  observational constraints are still consistent with our predictions (i.e.
  if we could still be missing a faint population of e.g. heavily obscured
  AGN).  We found that our prediction for the XRB contribution from
  luminosities limited to the range of overlap between observation and
  simulation provides an excess intensity of $1.85 \times 10^{-12} \rm{erg \: s^{-1}
  cm^{-2} deg^{-2}}$ in the 2-8 keV band (assuming photon index of 1.8).  This
  is below the 2-8 keV unresolved background of $I_{2-8 {\rm keV, unres}} = 3.4 \pm
  1.7 \times 10^{-12} \rm{erg \: s^{-1} cm^{-2} deg^{-2}}$
  \citep{2006ApJ...645...95H}.  If we take into account the total
  contribution from the whole population in the simulations (well below the
  faintest sources currently observed but still assuming the same X-ray
  spectrum) we are in excess of the unresolved background by almost a factor
  of 2.  To further illustrate and elucidate this issue, in Fig.
  \ref{BackgroundContributionVsLum} we plot the differential contribution to
  the 2-10 keV background from the simulations and the observations in several
  redshift bins (the filled curves are the regions bounded by the predictions
  from \citet{2003ApJ...598..886U}, \citet{2005ApJ...635..864L} and
  \citet{2009A&A...493...55E}). This shows again that the excess in our
  predictions is caused by the contribution from low-luminosity
  sources ($L < 10^{43} \rm{erg \: s^{-1}}$) and originates mostly at $z \ge 2$. This supports the
  idea that the faintest sources at high redshift are problematic in our
  predictions. Figs. 3 and 7 do indeed show that above z=2 this population is
  dominated by the lowest mass black holes (as opposed to lower redshifts
  where a more significant fraction of high mass black holes have
  'turned-off') and those that are likely to suffer more strongly from lack of
  resolution. For illustration, in Fig. \ref{LuminosityFunc}, at z = 2, in the hard X-ray band we
  show how the predictions look when only $M_{\rm BH} > 10^{7.5} M_{\odot}$ are
  plotted. The excess in our prediction is eliminated and the lowest
  luminosity end of the LF is now in good agreement with all the
  observations.

  Note also that we have used a redshift-independent correction to convert the
  simulations' bolometric luminosity to the hard X-ray band to compare with
  observations. Direct determinations of bolometric the correction as a function of
  redshift are not yet available and this may further bias our results 
  at $z\ge 2$.

\section{Conclusions}

Here we study the luminosity function and its evolution for populations of
quasars extracted from full cosmological hydrodynamical simulations which
include direct modelling for the growth of black holes. Noting that our
simulations (due to limitations on the volumes probed) can only be used to
study the faint-end of the QLF, we summarize our main results as follows:

\begin{itemize}
\item Consistent with the complex light-curves and various phases of activity
  that black holes undergo through their cosmic history, we have shown that
  there is a significant spread in luminosities for a given black hole mass,
  and in turn that different black hole masses contribute to the same regions
  of the QLF.

\item At low redshift ($z < 2$), the low luminosity ranges (below $10^9
  L_{\odot}$) of the QLF are dominated by black holes below $10^7 M_{\odot}$,
  while the luminosities above $10^9 L_{\odot}$ contain comparable
  contributions from both low and high masses.  At high redshift the
  majority of black holes are below $10^7 M_{\odot}$, and thus the entire
  QLF is dominated by low black hole mass sources.

\item We have shown that our predictions for the faint-end of the QLF agree
  remarkably well with observations at $z \le 1$, but produce steeper slopes
  than implied by current constraints for the hard X-ray band at redshifts $z
  = 2$ and 3. 

\item Taking into account a possible transition to low radiative
    efficiency accretion modes for low accretion rate sources tends to flatten
    the QLF at low redshifts but this does not affect significantly any of our
    results.

\item The exact form of the bolometric correction has a significant
    effect on our predictions. In particular, when comparing to a fixed
    correction, the empirically determined (Eq. 1) luminosity dependence leads
    to a larger QLF magnitude in the X-ray band.  Note also that in
    addition, no constraints are currently available on the redshift
    dependence of the correction.

\item The evolution of the comoving number density is in agreement with current
  constraints for the luminosity ranges above $10^{43} \rm{erg \: s^{-1}}$.  Agreement for
  luminosities below $10^{43} \rm{erg \: s^{-1}}$ is significantly worse, but the more
  limited observational data at these ranges combined with the dominance of
  black holes with $M_{\rm BH} < 10^7 M_{\odot}$, which are less-well resolved in
  our simulation makes these results less meaningful.

\item The luminosity density evolution predicts a peak luminosity density at
  $z =2.5$, with comparable contributions from different luminosity bins.

\item Based on the slope of the faint-end QLF, the luminosity density
  evolution and a moderate excess in the unresolved X-ray background, it
  appears that our simulations are overproducing low luminosity sources,
  particularly at intermediate redshifts. We have shown however that the
  higher resolution simulations produce fewer low-luminosity black holes, which makes
  it likely that this overproduction is dominated by resolution effects.
  Additionally, our results are most accurate at low redshift, when the high
  mass (and thus least likely to be affected by resolution limits) sources are most
  important, further suggesting that our overproduction of low luminosity
  sources is dominated by resolution effects.

  Overall our results support the interpretation of the faint-end luminosity
  function put forward by Hopkins et al. In upcoming work we will compare
  detailed characteristics of black hole lightcurves in our simulations and
  compare their instantaneous luminosities to their peak luminosities, so as
  to determine more precisely if the faint-end slope is dominated by quasars
  radiating below their peak, or by quasars with faint peak luminosities, as
  previous models assumed. It may be possible (although currently infeasible
  due to technological constraints) to run larger volume simulations at
  similar or higher resolution to increase the statistics at the bright-end of
  the luminosity function and further investigate the rapid dropoff in
  comoving number density at $z < 1$ found in the observational data.

\end{itemize}

\section*{Acknowledgments}

We would like to thank Debora Sijacki and the referee for their detailed comments and
suggestions which helped significantly improve this paper.

This work was supported by the National Science Foundation, NSF Petapps,
OCI-079212 and NSF AST-0607819.  The simulations were carried out at the NSF
Teragrid Pittsburgh Supercomputing Center (PSC). 


\label{lastpage}

 \bibliographystyle{mn2e}	
 \bibliography{astrobibl}	

\begin{thebibliography}{83}
\expandafter\ifx\csname natexlab\endcsname\relax\def\natexlab#1{#1}\fi

\bibitem[{Barger} et~al.(2003{\natexlab{a}}){Barger}, {Cowie}, {Capak}
  et~al.]{2003AJ....126..632B}
{Barger} A.~J., {Cowie} L.~L., {Capak} P., et~al., 2003{\natexlab{a}}, AJ, 126,
  632

\bibitem[{Barger} et~al.(2003{\natexlab{b}}){Barger}, {Cowie}, {Capak}
  et~al.]{2003ApJ...584L..61B}
{Barger} A.~J., {Cowie} L.~L., {Capak} P., et~al., 2003{\natexlab{b}}, \apjl,
  584, L61

\bibitem[{Barger} et~al.(2005){Barger}, {Cowie}, {Mushotzky}
  et~al.]{2005AJ....129..578B}
{Barger} A.~J., {Cowie} L.~L., {Mushotzky} R.~F., et~al., 2005, AJ, 129, 578

\bibitem[{Begelman} et~al.(2006){Begelman}, {Volonteri} \&
  {Rees}]{2006MNRAS.370..289B}
{Begelman} M.~C., {Volonteri} M., {Rees} M.~J., 2006, \mnras, 370, 289

\bibitem[{Bondi}(1952)]{1952MNRAS.112..195B}
{Bondi} H., 1952, \mnras, 112, 195

\bibitem[{Bondi} \& {Hoyle}(1944)]{1944MNRAS.104..273B}
{Bondi} H., {Hoyle} F., 1944, \mnras, 104, 273

\bibitem[{Bonoli} et~al.(2009){Bonoli}, {Marulli}, {Springel}, {White},
  {Branchini} \& {Moscardini}]{Bonoli2009}
{Bonoli} S., {Marulli} F., {Springel} V., {White} S.~D.~M., {Branchini} E.,
  {Moscardini} L., 2009, \mnras,  606

\bibitem[{Bower} et~al.(2006){Bower}, {Benson}, {Malbon}
  et~al.]{2006MNRAS.370..645B}
{Bower} R.~G., {Benson} A.~J., {Malbon} R., et~al., 2006, \mnras, 370, 645

\bibitem[{Boyle} et~al.(2000){Boyle}, {Shanks}, {Croom}
  et~al.]{2000MNRAS.317.1014B}
{Boyle} B.~J., {Shanks} T., {Croom} S.~M., et~al., 2000, \mnras, 317, 1014

\bibitem[{Brown} et~al.(2006){Brown}, {Brand}, {Dey}
  et~al.]{2006ApJ...638...88B}
{Brown} M.~J.~I., {Brand} K., {Dey} A., et~al., 2006, \apj, 638, 88

\bibitem[{Burkert} \& {Silk}(2001)]{2001ApJ...554L.151B}
{Burkert} A., {Silk} J., 2001, \apjl, 554, L151

\bibitem[{Churazov} et~al.(2005){Churazov}, {Sazonov}, {Sunyaev}, {Forman},
  {Jones} \& {B{\"o}hringer}]{2005MNRAS.363L..91C}
{Churazov} E., {Sazonov} S., {Sunyaev} R., {Forman} W., {Jones} C.,
  {B{\"o}hringer} H., 2005, \mnras, 363, L91

\bibitem[{Ciotti} \& {Ostriker}(2007)]{2007ApJ...665.1038C}
{Ciotti} L., {Ostriker} J.~P., 2007, \apj, 665, 1038

\bibitem[{Cirasuolo} et~al.(2005){Cirasuolo}, {Magliocchetti} \&
  {Celotti}]{2005MNRAS.357.1267C}
{Cirasuolo} M., {Magliocchetti} M., {Celotti} A., 2005, \mnras, 357, 1267

\bibitem[{Cowie} et~al.(2003){Cowie}, {Barger}, {Bautz}, {Brandt} \&
  {Garmire}]{2003ApJ...584L..57C}
{Cowie} L.~L., {Barger} A.~J., {Bautz} M.~W., {Brandt} W.~N., {Garmire} G.~P.,
  2003, \apjl, 584, L57

\bibitem[{Cristiani} et~al.(2004){Cristiani}, {Alexander}, {Bauer}
  et~al.]{2004ApJ...600L.119C}
{Cristiani} S., {Alexander} D.~M., {Bauer} F., et~al., 2004, \apjl, 600, L119

\bibitem[{Croom} et~al.(2004){Croom}, {Smith}, {Boyle}
  et~al.]{2004MNRAS.349.1397C}
{Croom} S.~M., {Smith} R.~J., {Boyle} B.~J., et~al., 2004, \mnras, 349, 1397

\bibitem[{Croton} et~al.(2006){Croton}, {Springel}, {White}
  et~al.]{2006MNRAS.365...11C}
{Croton} D.~J., {Springel} V., {White} S.~D.~M., et~al., 2006, \mnras, 365, 11

\bibitem[{Di Matteo} et~al.(2008){Di Matteo}, {Colberg}, {Springel},
  {Hernquist} \& {Sijacki}]{2008ApJ...676...33D}
{Di Matteo} T., {Colberg} J., {Springel} V., {Hernquist} L., {Sijacki} D.,
  2008, \apj, 676, 33

\bibitem[{Di Matteo} et~al.(2005){Di Matteo}, {Springel} \&
  {Hernquist}]{2005Natur...433..604D}
{Di Matteo} T., {Springel} V., {Hernquist} L., 2005, Nature, 433, 604

\bibitem[{Ebrero} et~al.(2009){Ebrero}, {Carrera}, {Page}
  et~al.]{2009A&A...493...55E}
{Ebrero} J., {Carrera} F.~J., {Page} M.~J., et~al., 2009, A\&A, 493, 55

\bibitem[{Fan} et~al.(2004){Fan}, {Hennawi}, {Richards}
  et~al.]{2004AJ....128..515F}
{Fan} X., {Hennawi} J.~F., {Richards} G.~T., et~al., 2004, AJ, 128, 515

\bibitem[{Fan} et~al.(2001{\natexlab{a}}){Fan}, {Narayanan}, {Lupton}
  et~al.]{2001AJ....122.2833F}
{Fan} X., {Narayanan} V.~K., {Lupton} R.~H., et~al., 2001{\natexlab{a}}, AJ,
  122, 2833

\bibitem[{Fan} et~al.(2001{\natexlab{b}}){Fan}, {Strauss}, {Schneider}
  et~al.]{2001AJ....121...54F}
{Fan} X., {Strauss} M.~A., {Schneider} D.~P., et~al., 2001{\natexlab{b}}, AJ,
  121, 54

\bibitem[{Fan} et~al.(2003){Fan}, {Strauss}, {Schneider}
  et~al.]{2003AJ....125.1649F}
{Fan} X., {Strauss} M.~A., {Schneider} D.~P., et~al., 2003, AJ, 125, 1649

\bibitem[{Ferrarese} \& {Merritt}(2000)]{2000ApJ...539L...9F}
{Ferrarese} L., {Merritt} D., 2000, \apjl, 539, L9

\bibitem[{Fiore} et~al.(2003){Fiore}, {Brusa}, {Cocchia}
  et~al.]{2003A&A...409...79F}
{Fiore} F., {Brusa} M., {Cocchia} F., et~al., 2003, A\&A, 409, 79

\bibitem[{Gebhardt} et~al.(2000){Gebhardt}, {Bender}, {Bower}
  et~al.]{2000ApJ...539L..13G}
{Gebhardt} K., {Bender} R., {Bower} G., et~al., 2000, \apjl, 539, L13

\bibitem[{Graham} \& {Driver}(2007)]{2007ApJ...655...77G}
{Graham} A.~W., {Driver} S.~P., 2007, \apj, 655, 77

\bibitem[{Granato} et~al.(2004){Granato}, {De Zotti}, {Silva}, {Bressan} \&
  {Danese}]{2004ApJ...600..580G}
{Granato} G.~L., {De Zotti} G., {Silva} L., {Bressan} A., {Danese} L., 2004,
  \apj, 600, 580

\bibitem[{Haiman} et~al.(2004){Haiman}, {Quataert} \& {Bower}]{Haiman2004}
{Haiman} Z., {Quataert} E., {Bower} G.~C., 2004, \apj, 612, 698

\bibitem[{Hasinger} et~al.(2005){Hasinger}, {Miyaji} \&
  {Schmidt}]{2005A&A...441..417H}
{Hasinger} G., {Miyaji} T., {Schmidt} M., 2005, A\&A, 441, 417

\bibitem[{Hickox} \& {Markevitch}(2006)]{2006ApJ...645...95H}
{Hickox} R.~C., {Markevitch} M., 2006, \apj, 645, 95

\bibitem[{Hopkins} et~al.(2005{\natexlab{a}}){Hopkins}, {Hernquist}, {Cox}
  et~al.]{2005ApJ...630..705H}
{Hopkins} P.~F., {Hernquist} L., {Cox} T.~J., et~al., 2005{\natexlab{a}}, \apj,
  630, 705

\bibitem[{Hopkins} et~al.(2005{\natexlab{b}}){Hopkins}, {Hernquist}, {Cox}, {Di
  Matteo}, {Robertson} \& {Springel}]{2005ApJ...630..716H}
{Hopkins} P.~F., {Hernquist} L., {Cox} T.~J., {Di Matteo} T., {Robertson} B.,
  {Springel} V., 2005{\natexlab{b}}, \apj, 630, 716

\bibitem[{Hopkins} et~al.(2006{\natexlab{a}}){Hopkins}, {Hernquist}, {Cox}, {Di
  Matteo}, {Robertson} \& {Springel}]{2006ApJS..163....1H}
{Hopkins} P.~F., {Hernquist} L., {Cox} T.~J., {Di Matteo} T., {Robertson} B.,
  {Springel} V., 2006{\natexlab{a}}, \apjs, 163, 1

\bibitem[{Hopkins} et~al.(2006{\natexlab{b}}){Hopkins}, {Hernquist}, {Cox},
  {Robertson}, {Di Matteo} \& {Springel}]{2006ApJ...639..700H}
{Hopkins} P.~F., {Hernquist} L., {Cox} T.~J., {Robertson} B., {Di Matteo} T.,
  {Springel} V., 2006{\natexlab{b}}, \apj, 639, 700

\bibitem[{Hopkins} et~al.(2007{\natexlab{a}}){Hopkins}, {Hernquist}, {Cox},
  {Robertson} \& {Krause}]{2007ApJ...669...45H}
{Hopkins} P.~F., {Hernquist} L., {Cox} T.~J., {Robertson} B., {Krause} E.,
  2007{\natexlab{a}}, \apj, 669, 45

\bibitem[{Hopkins} et~al.(2005{\natexlab{c}}){Hopkins}, {Hernquist}, {Martini}
  et~al.]{2005ApJ...625L..71H}
{Hopkins} P.~F., {Hernquist} L., {Martini} P., et~al., 2005{\natexlab{c}},
  \apjl, 625, L71

\bibitem[{Hopkins} et~al.(2007{\natexlab{b}}){Hopkins}, {Richards} \&
  {Hernquist}]{2007ApJ...654..731H}
{Hopkins} P.~F., {Richards} G.~T., {Hernquist} L., 2007{\natexlab{b}}, \apj,
  654, 731

\bibitem[{Hoyle} \& {Lyttleton}(1939)]{1939PCPS...35..405H}
{Hoyle} F., {Lyttleton} R.~A., 1939, in { Proceedings of the Cambridge
  Philosophical Society\/}, vol.~35 of { Proceedings of the Cambridge
  Philosophical Society\/},  405

\bibitem[{Hunt} et~al.(2004){Hunt}, {Steidel}, {Adelberger} \&
  {Shapley}]{2004ApJ...605..625H}
{Hunt} M.~P., {Steidel} C.~C., {Adelberger} K.~L., {Shapley} A.~E., 2004, \apj,
  605, 625

\bibitem[{Kauffmann} \& {Haehnelt}(2000)]{2000MNRAS.311..576K}
{Kauffmann} G., {Haehnelt} M., 2000, \mnras, 311, 576

\bibitem[{Kawata} \& {Gibson}(2005)]{2005MNRAS.358L..16K}
{Kawata} D., {Gibson} B.~K., 2005, \mnras, 358, L16

\bibitem[{Kennefick} et~al.(1995){Kennefick}, {Djorgovski} \& {de
  Carvalho}]{1995AJ....110.2553K}
{Kennefick} J.~D., {Djorgovski} S.~G., {de Carvalho} R.~R., 1995, AJ, 110, 2553

\bibitem[{Kormendy} \& {Richstone}(1995)]{1995ARA&A..33..581K}
{Kormendy} J., {Richstone} D., 1995, ARA\&A, 33, 581

\bibitem[{La Franca} et~al.(2005){La Franca}, {Fiore}, {Comastri}
  et~al.]{2005ApJ...635..864L}
{La Franca} F., {Fiore} F., {Comastri} A., et~al., 2005, \apj, 635, 864

\bibitem[{La Franca} et~al.(2002){La Franca}, {Fiore}, {Vignali}
  et~al.]{2002ApJ...570..100L}
{La Franca} F., {Fiore} F., {Vignali} C., et~al., 2002, \apj, 570, 100

\bibitem[{Lewis} et~al.(2002){Lewis}, {Cannon}, {Taylor}
  et~al.]{2002MNRAS.333..279L}
{Lewis} I.~J., {Cannon} R.~D., {Taylor} K., et~al., 2002, \mnras, 333, 279

\bibitem[{Magorrian} et~al.(1998){Magorrian}, {Tremaine}, {Richstone}
  et~al.]{1998AJ....115.2285M}
{Magorrian} J., {Tremaine} S., {Richstone} D., et~al., 1998, AJ, 115, 2285

\bibitem[{Malbon} et~al.(2007){Malbon}, {Baugh}, {Frenk} \&
  {Lacey}]{2007MNRAS.382.1394M}
{Malbon} R.~K., {Baugh} C.~M., {Frenk} C.~S., {Lacey} C.~G., 2007, \mnras, 382,
  1394

\bibitem[{Marconi} et~al.(2004){Marconi}, {Risaliti}, {Gilli}, {Hunt},
  {Maiolino} \& {Salvati}]{2004MNRAS.351..169M}
{Marconi} A., {Risaliti} G., {Gilli} R., {Hunt} L.~K., {Maiolino} R., {Salvati}
  M., 2004, \mnras, 351, 169

\bibitem[{Marulli} et~al.(2008){Marulli}, {Bonoli}, {Branchini}, {Moscardini}
  \& {Springel}]{Marulli2008}
{Marulli} F., {Bonoli} S., {Branchini} E., {Moscardini} L., {Springel} V.,
  2008, \mnras, 385, 1846

\bibitem[{Matute} et~al.(2006){Matute}, {La Franca}, {Pozzi}, {Gruppioni},
  {Lari} \& {Zamorani}]{2006A&A...451..443M}
{Matute} I., {La Franca} F., {Pozzi} F., {Gruppioni} C., {Lari} C., {Zamorani}
  G., 2006, A\&A, 451, 443

\bibitem[{Miyaji} et~al.(2001){Miyaji}, {Hasinger} \&
  {Schmidt}]{2001A&A...369...49M}
{Miyaji} T., {Hasinger} G., {Schmidt} M., 2001, A\&A, 369, 49

\bibitem[{Moretti} et~al.(2003){Moretti}, {Campana}, {Lazzati} \&
  {Tagliaferri}]{Moretti2003}
{Moretti} A., {Campana} S., {Lazzati} D., {Tagliaferri} G., 2003, \apj, 588,
  696

\bibitem[{Nandra} et~al.(2005){Nandra}, {Laird} \&
  {Steidel}]{2005MNRAS.360L..39N}
{Nandra} K., {Laird} E.~S., {Steidel} C.~C., 2005, \mnras, 360, L39

\bibitem[{Narayan}(2005)]{Narayan2005}
{Narayan} R., 2005, AP\&SS, 300, 177

\bibitem[{Okamoto} et~al.(2008){Okamoto}, {Nemmen} \&
  {Bower}]{2008MNRAS.385..161O}
{Okamoto} T., {Nemmen} R.~S., {Bower} R.~G., 2008, \mnras, 385, 161

\bibitem[{Page} et~al.(1997){Page}, {Mason}, {McHardy}, {Jones} \&
  {Carrera}]{1997MNRAS.291..324P}
{Page} M.~J., {Mason} K.~O., {McHardy} I.~M., {Jones} L.~R., {Carrera} F.~J.,
  1997, \mnras, 291, 324

\bibitem[{Peacock}(1999)]{1999coph.book.....P}
{Peacock} J.~A., 1999, {Cosmological Physics}, Cosmological Physics, by John
  A.~Peacock, pp.~704.~ISBN 052141072X.~Cambridge, UK: Cambridge University
  Press, January 1999.

\bibitem[{Quataert} \& {Narayan}(1999)]{QuataertNarayan1999}
{Quataert} E., {Narayan} R., 1999, \apj, 520, 298

\bibitem[{Richards} et~al.(2005){Richards}, {Croom}, {Anderson}
  et~al.]{2005MNRAS.360..839R}
{Richards} G.~T., {Croom} S.~M., {Anderson} S.~F., et~al., 2005, \mnras, 360,
  839

\bibitem[{Richards} et~al.(2006){Richards}, {Strauss}, {Fan}
  et~al.]{2006AJ....131.2766R}
{Richards} G.~T., {Strauss} M.~A., {Fan} X., et~al., 2006, AJ, 131, 2766

\bibitem[{Sazonov} et~al.(2004){Sazonov}, {Ostriker} \&
  {Sunyaev}]{2004MNRAS.347..144S}
{Sazonov} S.~Y., {Ostriker} J.~P., {Sunyaev} R.~A., 2004, \mnras, 347, 144

\bibitem[{Schmidt} et~al.(1995){Schmidt}, {Schneider} \&
  {Gunn}]{1995AJ....110...68S}
{Schmidt} M., {Schneider} D.~P., {Gunn} J.~E., 1995, AJ, 110, 68

\bibitem[{Shakura} \& {Sunyaev}(1973)]{1973A&A....24..337S}
{Shakura} N.~I., {Sunyaev} R.~A., 1973, A\&A, 24, 337

\bibitem[{Siana} et~al.(2006){Siana}, {Polletta}, {Smith}
  et~al.]{2006astro.ph..4373S}
{Siana} B., {Polletta} M., {Smith} H.~E., et~al., 2006, ArXiv Astrophysics
  e-prints:astro-ph/0604373

\bibitem[{Sijacki} et~al.(2007){Sijacki}, {Springel}, {di Matteo} \&
  {Hernquist}]{2007MNRAS.380..877S}
{Sijacki} D., {Springel} V., {di Matteo} T., {Hernquist} L., 2007, \mnras, 380,
  877

\bibitem[{Silverman} et~al.(2005){Silverman}, {Green}, {Barkhouse}
  et~al.]{2005ApJ...618..123S}
{Silverman} J.~D., {Green} P.~J., {Barkhouse} W.~A., et~al., 2005, \apj, 618,
  123

\bibitem[{Silverman} et~al.(2008){Silverman}, {Green}, {Barkhouse}
  et~al.]{2008ApJ...679..118S}
{Silverman} J.~D., {Green} P.~J., {Barkhouse} W.~A., et~al., 2008, \apj, 679,
  118

\bibitem[{Springel} et~al.(2005{\natexlab{a}}){Springel}, {Di Matteo} \&
  {Hernquist}]{2005ApJ...620L..79S}
{Springel} V., {Di Matteo} T., {Hernquist} L., 2005{\natexlab{a}}, \apjl, 620,
  L79

\bibitem[{Springel} et~al.(2005{\natexlab{b}}){Springel}, {Di Matteo} \&
  {Hernquist}]{2005MNRAS.361..776S}
{Springel} V., {Di Matteo} T., {Hernquist} L., 2005{\natexlab{b}}, \mnras, 361,
  776

\bibitem[{Tremaine} et~al.(2002){Tremaine}, {Gebhardt}, {Bender}
  et~al.]{2002ApJ...574..740T}
{Tremaine} S., {Gebhardt} K., {Bender} R., et~al., 2002, \apj, 574, 740

\bibitem[{Treu} et~al.(2007){Treu}, {Woo}, {Malkan} \&
  {Blandford}]{2007ApJ...667..117T}
{Treu} T., {Woo} J.-H., {Malkan} M.~A., {Blandford} R.~D., 2007, \apj, 667, 117

\bibitem[{Ueda} et~al.(2003){Ueda}, {Akiyama}, {Ohta} \&
  {Miyaji}]{2003ApJ...598..886U}
{Ueda} Y., {Akiyama} M., {Ohta} K., {Miyaji} T., 2003, \apj, 598, 886

\bibitem[{Volonteri} et~al.(2003){Volonteri}, {Haardt} \&
  {Madau}]{Volonteri2003}
{Volonteri} M., {Haardt} F., {Madau} P., 2003, \apj, 582, 559

\bibitem[{Wall} et~al.(2005){Wall}, {Jackson}, {Shaver}, {Hook} \&
  {Kellermann}]{2005A&A...434..133W}
{Wall} J.~V., {Jackson} C.~A., {Shaver} P.~A., {Hook} I.~M., {Kellermann}
  K.~I., 2005, A\&A, 434, 133

\bibitem[{Wolf} et~al.(2003){Wolf}, {Wisotzki}, {Borch}, {Dye}, {Kleinheinrich}
  \& {Meisenheimer}]{2003A&A...408..499W}
{Wolf} C., {Wisotzki} L., {Borch} A., {Dye} S., {Kleinheinrich} M.,
  {Meisenheimer} K., 2003, A\&A, 408, 499

\bibitem[{Wyithe} \& {Loeb}(2003)]{2003ApJ...595..614W}
{Wyithe} J.~S.~B., {Loeb} A., 2003, \apj, 595, 614

\bibitem[{Yencho} et~al.(2009){Yencho}, {Barger}, {Trouille} \&
  {Winter}]{Yencho2009}
{Yencho} B., {Barger} A.~J., {Trouille} L., {Winter} L.~M., 2009, \apj, 698,
  380

\bibitem[{York} et~al.(2000){York}, {Adelman}, {Anderson}
  et~al.]{2000AJ....120.1579Y}
{York} D.~G., {Adelman} J., {Anderson} Jr. J.~E., et~al., 2000, AJ, 120, 1579

\bibitem[{Yuan} \& {Narayan}(2004)]{YuanNarayan2004}
{Yuan} F., {Narayan} R., 2004, \apj, 612, 724

\end{thebibliography}

\end{document}